\documentclass{aa}
\usepackage{psfig} 

\def\ros{{\sl ROSAT }}

\def\G{$\Gamma_{\rm x}$ }

\def\fss{\hbox{$.\!\!^{\rm s}$}}        
\def\amin{\ifmmode ^{\prime}\else$^{\prime}$\fi}
\def\asec{\ifmmode ^{\prime\prime}\else$^{\prime\prime}$\fi}
\def\h{$^{\rm h}$}\def\m{$^{\rm m}$}

\def\etal{{et\,al.}}
\def\approxlt{\mathrel{\hbox{\rlap{\lower.55ex \hbox {$\sim$}}
        \kern-.3em \raise.4ex \hbox{$<$}}}}
\def\approxgt{\mathrel{\hbox{\rlap{\lower.55ex \hbox {$\sim$}}
        \kern-.3em \raise.4ex \hbox{$>$}}}}

\begin{document}

  \thesaurus{03         
             (02.01.2; 
               11.01.2;  
               11.09.1;  
               11.14.1;  
               13.25.2)  
}

 \title{Discovery of a giant and luminous X-ray outburst from  \\
    the optically inactive galaxy pair RX\,J1242.6--1119\thanks{Partly based on observations 
    obtained at the European Southern Observatory, La Silla, Chile.} }  

  \author{ Stefanie Komossa\inst{1}, Jochen Greiner \inst{2}}

  \offprints{St. Komossa, skomossa@mpe.mpg.de}
  \institute{
      Max-Planck-Institut f\"ur extraterrestrische Physik,
         Postfach 1603, D-85740 Garching, Germany
     \and
       Astrophysikalisches Institut Potsdam, An der Sternwarte 16, 
        D-14482 Potsdam, Germany}
\date{Received: 30 July 1999; accepted: 17 August 1999 }

   \maketitle
\markboth{St.~Komossa and J. Greiner: A giant X-ray outburst from the galaxy pair  
          RX\,J1242.6-1119}
{St.~Komossa and J. Greiner: A giant X-ray outburst from the galaxy pair RX\,J1242.6-1119}

   \begin{abstract}
We report the discovery of large-amplitude X-ray variability 
from the direction of the previously unknown, optically inactive 
galaxy pair RX\,J1242.6-1119.
The X-ray source shows variability by a factor $\approxgt$ 20 between
the \ros all-sky survey and a later pointed observation separated by
$\sim$1.5 yr. Its spectrum is extremely soft with photon index
\G $\simeq -5$, among the steepest ever observed among galaxies.
Based on the redshift derived from the optical spectra, $z$=0.05,
the source's intrinsic luminosity is large, $L_{\rm x} \approxgt$ 9\,10$^{43}$ 
erg\,s$^{-1}$.  
Surprisingly, the optical spectra of both galaxies are characterized by absorption
lines and do not show signs of (Seyfert) activity. 
This makes RX\,J1242-11 the third candidate for giant-amplitude 
variability in an otherwise non-active galaxy, the first two 
being NGC\,4552 (in the UV; Renzini et al. 1995) and NGC\,5905
(in X-rays; Bade et al. 1996, Komossa \& Bade 1999).
Several mechanisms to explain this unexpected and peculiar
behavior are investigated. 
The most likely one seems to be an accretion event 
onto an otherwise dormant supermassive black hole (SMBH),
e.g., by a tidal disruption event.  

   \end{abstract}
%
\section{Introduction}

Giant X-ray outbursts among galaxies, even the active ones, are rare.
`Normal' galaxies, starbursts, and most LINERs show constant
soft X-ray emission (e.g., Fabbiano 1989, Vogler \& Pietsch 1999, Komossa et al. 1999)
and X-ray luminosities of typically 10$^{38-40}$ erg\,s$^{-1}$. 
Some off-nuclear, X-ray-bright point sources, several of them variable,
have been detected recently but they seldom exceed 10$^{39}$ erg\,s$^{-1}$ 
(e.g., Immler et al. 1998, Komossa \& Schulz 1998, and references therein).  
In contrast, many active galactic nuclei 
(AGN; $L_{\rm x} \approxgt 10^{42}$ erg\,s$^{-1}$) are known to be 
X-ray variable by typically a factor 2--3 (e.g., Mushotzky et al. 1993). 
Outbursts with amplitudes exceeding a factor $\sim$10-20 are very rare
even among AGN, though.
Therefore, it was quite surprising when giant X-ray outbursts
(factors $\approxgt$100 in countrate) 
from the two optically `rather' inactive galaxies
IC\,3599 (Brandt et al. 1995, Grupe et al. 1995)
and NGC\,5905 (Bade et al. 1996, Komossa \& Bade 1999) 
were discovered. Both outbursters were characterized
by very soft X-ray spectra and reached high outburst luminosities.  

The \ros all-sky survey (RASS; Voges et al. 1996) provides an excellent data base
to search for 
further cases of giant X-ray variability. 
Such outbursts provide a powerful tool to
probe for the existence of SMBHs in nearby galaxies
and to study 
the physics of accretion events.   
In particular, Rees (1988,1989,1990)
proposed to use the UV-X-ray flares expected
from the tidal disruption events of stars swallowed
by SMBHs to detect these SMBHs in
nearby, {\em non-active} galaxies.

RXJ1242-11, serendipituously located in the field-of-view
of a \ros PSPC pointing, was originally selected for optical 
follow-up observations due to its very soft X-ray spectrum 
in the course of the extension of the identification program of 
supersoft X-ray sources from the RASS (Greiner 1996) to pointed
observations of nearby galaxies. 
We report here the optical and X-ray properties of this 
previously unknown source (Sect. 2) and discuss
scenarios to account for its very peculiar variability behavior (Sect. 3).  
Luminosities given below are calculated assuming $H_0=50$ km/s/Mpc.

\section{Data reduction and analysis} 

\subsection{Optical observations}

Spectra and images were acquired at the 1.5m Danish telescope
at ESO/La Silla on January 24 and 26, 1999, equipped with DFOSC. 
A grism with 300 grooves per mm was used yielding a dispersion of 
3.3 \AA/pixel on the 2052*2052 backside illuminated LORAL/LESSER CCD. 
With a 1\farcs5 slit the FWHM resolution is 12 \AA. 
Exposure times were 900 and 2200 sec, respectively, and the spectra were
debiased, flatfielded and calibrated (with the standard star GD 108)
using standard MIDAS procedures.

\vskip0.2cm
\noindent {\em Results: Image.} 
The image shown in Fig. 1 is the sum of 18 acquisition images (effective
exposure time of 7 min) and reveals a close pair of galaxies within
the X-ray error circle. These galaxies were not resolved on the POSS,
thus leading to entries in the USNO-A1.0 catalogue (Monet \etal\ 1998):
U0750\_07951028 (the optically brighter one of V=14$^{\rm m}$) and 
U0750\_07951000 (V=16$^{\rm m}$). 
They are probably interacting,
as indicated by a `light bridge' between the two. 
Further objects visible in the image are fainter than V=21.8 mag.
The optical coordinates (J2000) of the two galaxies are
$\alpha$=12\h 42\m 38\fss5, $\delta$=--11\degr 19\amin 21\asec (U0750\_07951028) 
and 
$\alpha$=12\h 42\m 38\fss2, $\delta$=--11\degr 19\amin 15\asec (U0750\_07951000).

\begin{figure}
\vbox{\psfig{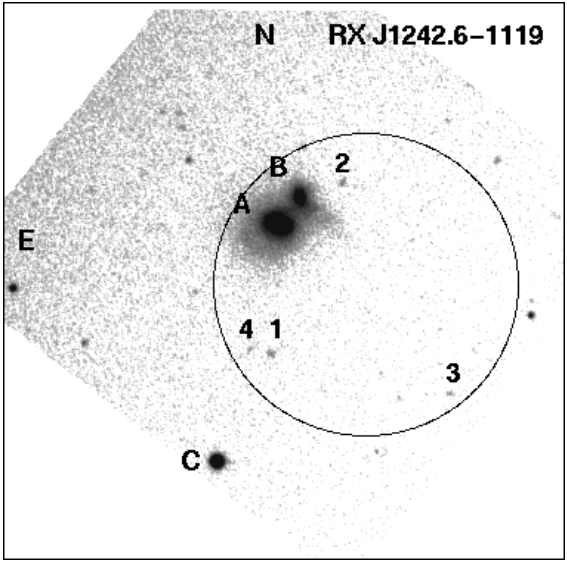}}\par
\caption[over]{A 2\farcm5$\times$2\farcm5 V band image of the region around 
   RX\,J1242.6--1119. The circle marks the X-ray position uncertainty
 of 40\arcsec\ radius. ``A'' and ``B'' denote the galaxy pair, and numbers 1--4
  mark faint objects in the error circle, sorted according to brightness:
  21.8, 22.0, 22.0 and 22.4 mag, respectively (derived by comparison with USNO star
    U0750\_07951099 which is labeled ``C'').
} 
\label{over}
\end{figure} 

\vskip0.2cm
\noindent {\em Results: Spectra.}
The optical spectrum of the brighter of the two galaxies 
is characterized by strong absorption
lines of Na\,I 5175\AA, Mg\,I 5890\AA~and H$\beta$ (Fig. 2;
note that H$\alpha$ overlaps with the atmospheric B band).
Using the Mg\,I and Na\,I line we derive 
a redshift of $z=0.050\pm0.001$. 
The slope of the spectrum suggests either an elliptical or early spiral type.
The Balmer absorption lines indicate the presence of A stars. 
Both, these deep Balmer lines as well as the strong drop of the
flux beyond the Ca\,II H/K break argue against a classification as
BL Lac object.             
No AGN-like forbidden emission lines, like [OIII]$\lambda$5007, are detected. 

The spectrum of the second, optically weaker galaxy looks very similar to the
first one (Fig. 2, lower panel). Using the same absorption lines as above, we derive  
the same redshift of $z=0.05$. Although, again, this spectrum is not
like those of AGN, we note that there appears
to be some excess emission close to the location of the [OIII]$\lambda$5007
line.  

\begin{figure}
\psfig{file=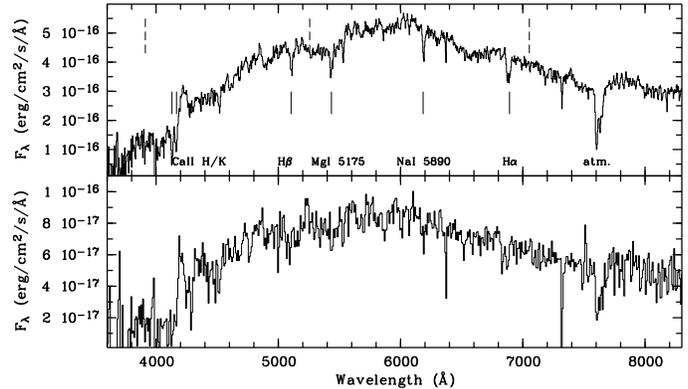,width=8.5cm}
 \vspace*{-0.4cm}
 \caption[spec]{ Spectra of the two galaxies with the brighter one rebinned by a factor of 
    4 to match the bin size with the spectral resolution (top panel) and the
    fainter one rebinned by a factor of 8 (bottom panel). Prominent absorption
    lines are marked, together with the expected locations of undetected
    emission lines of [OII]$\lambda$3727, [OIII]$\lambda$5007 and [SII]$\lambda$6717 
    (dashed lines, from left to 
    right). The feature near 4200\AA~which appears in both spectra is due to
    the improper subtraction of a large, spherical distortion in the 
    flatfield image. The absorption near 7600\AA~is atmospheric in origin. 
}
\label{spec}
\end{figure}

\subsection{X-ray observations}

The region of RX\,J1242-11 was observed twice with the 
\ros (Tr\"umper 1983, Briel et al. 1994) PSPC. 
First in December 1990 -- January 1991 during the RASS,
then serendipituously in a pointed observation of July 1992   
where RX\,J1242-11 is located 40\arcmin~off-axis. 
Details of the observations are given in Table 1.
The data reduction was carried out in a standard manner.
In brief,  
for the spectral and temporal analysis of the PSPC pointed data 
source photons were extracted
within a circular region of radius 400\arcsec~centered on the target source.
The background was determined in a source-free ring around
the target source and subtracted.
The data were corrected for vignetting
using the EXSAS software package (Zimmermann et al. 1994).
To carry out the spectral analysis source photons in
the amplitude channels 11-240 were binned
according to a constant signal/noise ratio of 5$\sigma$.

   \begin{table*}[ht]
     \caption{Log of the X-ray observations and results from spectral fits.
              $t_{\rm exp}$ gives the exposure time,
              $CR$ the countrate in the (0.1-2.4) keV band. \G is the photon index
              derived from the powerlaw spectral fit, $kT$ the black body
              temperature for the black body fit (see text for details). In both 
              cases,
              we fixed $N_{\rm H}$ = $N_{\rm Gal}$. Errors in \G and $kT$ 
              are quoted at 68\% confidence. 
              $L_{\rm x}$ gives the intrinsic luminosity in the
              (0.1--2.4) keV band.  
       }
     \label{obslog}
      \begin{tabular}{llrcccccc}
      \hline
        observ. & date & $t_{\rm exp}$ & $CR$ & X-ray coor. (J2000) & \G & 
$L_{\rm x, pl}$
              &  $kT_{\rm bb}$ & $L_{\rm x, bb}$  \\
             &        &  sec          & cts/s & RA ~~~~~ DEC &  & 10$^{43}$ erg/s 
               & keV & 10$^{43}$ erg/s   \\
      \hline
      \hline
  RASS  &  28-29/12/90\,+\,17/1/91  &  270 & $<$0.015 &  & &  & & \\
  point.$^{*}$ &  15-19/7/92 & 10567 & 0.3
      &  12\h42\m36\fss9 ~~ --11\degr19{$\arcmin$}35{\arcsec} &
         --5.1$\pm{0.9}$ & 35.5 & 0.06$\pm{0.01}$ & 8.8 \\
      \hline
  \end{tabular}

\noindent{$^{*}$ Observation identification number 600258p
}
   \end{table*}

\subsubsection {Temporal analysis} 

Whereas RX\,J1242-11 is {\em brighter} in X-rays than the galaxy on which the 
pointing was centered, M\,104, RX\,J1242-11 is {\em not detected at all} during 
the RASS.
In contrast, M\,104 is again 
clearly detected.
This immediately reveals large-amplitude variability of RX\,J1242-11.

We estimate a 3$\sigma$ upper limit countrate of 0.015 cts/s during 
the RASS observation.
In the pointed observation, the source is 
partially hidden behind the detector support rib structure. 
Not correcting for the partial shadowing effect, we derive a mean
source countrate of 0.125$\pm{0.005}$ cts/s.
Following the variation of the source's countrate over the
wobble paths, the countrate reaches maximal values of 0.3 cts/s
and we take this as the best estimate for the undisturbed source emission.    
Comparison with the countrate upper limit derived for the
RASS observation reveals variability by a factor larger than 20.

\subsubsection {Spectral analysis} 

Several spectral models were fit to the X-ray emission, 
starting with a powerlaw and cold absorption
fixed to the Galactic value in direction of RXJ1242-11,
$N_{\rm Gal} = 3.74\,10^{20}$ cm$^{-2}$ (Dickey \& Lockman 1990). 
We find \G=$-5.1\pm{0.9}$ ($\chi^2_{\rm red}$ = 1.5). 
No source photons are detected above $\sim$1\,KeV. 
Treating $N_{\rm H}$ as free parameter, 
the powerlaw becomes even steeper (\G $\simeq -9$ and 
$N_{\rm H} \simeq 3\,N_{\rm Gal}$, but both parameters are no longer well constrained). 
Alternatively, a black body was fit. 
This yields $kT_{\rm bb}$ = 0.06$\pm{0.01}$ keV for 
$N_{\rm H}=N_{\rm Gal}$ and gives an excellent 
fit ($\chi^2_{\rm red}$ = 0.7). We repeated all spectral fits using 
a second background region and find the same best-fit parameters
 within the errors.    
Results are summarized in Table 1. 

Using the powerlaw (black body) fit and $N_{\rm H}=N_{\rm Gal}$,
we derive an intrinsic X-ray luminosity of 
$L_{\rm x}$ = 35.5\,(8.8) $\times$ 10$^{43}$ erg\,s$^{-1}$. 
This is a lower limit on the actual peak luminosity
emitted during outburst, 
since we most likely have not caught the source exactly at maximum
light, since the spectrum may extend into the EUV, and since there
may be excess absorption along the line-of-sight.
Each of the three effects could easily boost $L_{\rm x}$ by an order
of magnitude.  
 
\section{Discussion}

In summary, the properties to be explained by any
outburst model are: (i) high outburst luminosity,
(ii) very soft X-ray spectrum, and (iii) optical
spectrum of a non-active galaxy.  
Below, we briefly compare with the few other cases of UV-X-ray
variability among non-active galaxies 
reported so far, and then discuss possible scenarios for RX\,J1242-11. 

\subsection {Previous observations of large-amplitude X-ray outbursts
   in optically non-active galaxies}

IC\,3599 (Brandt et al. 1995, Grupe et al. 1995)
and NGC\,5905 (Bade et al. 1996) both underwent giant X-ray outbursts.
The optical spectrum of IC\,3599 shows several indications
of activity even in X-ray quiescence (Komossa \& Bade 1999; KoBa99 hereafter)
which is not the case for NGC\,5905. Its spectrum is that of an HII galaxy.
Many outburst models were studied,  
and  tidal disruption of a star by
a SMBH (e.g., Rees 1988, 1990; Loeb \& Ulmer 1997) 
was tentatively favored as explanation.
Another strongly X-ray variable source,  
with an optical spectrum quite similar to RX\,J1242-11
will be presented by Grupe et al. (A\&A, subm).
It is also worthwhile to mention NGC\,3628 which showed
a {\em drop} by a factor 20 in \ros flux
(Dahlem et al. 1995).
In the UV spectral region, an outburst was detected from the elliptical
galaxy NGC\,4552, interpreted
by Renzini et al. (1995) as accretion event (the tidal stripping
of a star's atmosphere by a SMBH, or the accretion of a molecular cloud).

\subsection {Variability mechanisms}

In the discussion below, we assume that indeed an outburst occurred, 
instead
of a transient drop in luminosity. Whereas the outburst character was well 
evidenced
by the long-term X-ray lightcurve of NGC\,5905 (cf. Fig. 9 of KoBa99),
we only have two measurements for RXJ1242-11 so far, one in low-state,
one in high-state.  
Besides the analogy to the previous outbursting sources  
a further argument in favor of an outburst
is that the continuous X-ray emission of `normal' galaxies usually
does not exceed 10$^{39-41}$ erg\,s$^{-1}$ and is extended.  

Further, we note that the presence of a Galactic foreground
object is unlikely, since further optical sources
within the X-ray error circle are extremely weak, 
and RXJ\,1242-11 is at high Galactic latitude (b$_{\rm II}$=51\fdg5). 
Given the high $L_{\rm X}/L_{\rm opt}$ value, an ISM accreting neutron
star might come to mind,  
but the strong X-ray variability
would require an extreme ISM density gradient, thus leading us to reject
this possibility.

Many outburst scenarios were already discussed for the
case of NGC\,5905 (KoBa99). 
Several (variable stellar sources, like X-ray binaries, supernovae (SN),
or SN in dense medium) turned out to be very 
unlikely since they cannot fulfill the tight constraint set by the
huge outburst luminosity.
This similarly holds for RX\,J1242-11.
Further models (for details see the estimates and references
in KoBa99) are discussed in turn: 

Could we have witnessed the X-ray afterglow of a GRB~?
This is unlikely, since the ``on" timescale of several
days is much too long as compared to all known cases of GRB afterglows
(e.g.,  Greiner et al. 1999),
which quickly faded with a $t^{-1}$ law after detection.

The presence of a hidden Seyfert nucleus obscured
by a {\em dusty} warm absorber is in conflict with the very steep
observed X-ray spectrum.   

Remaining scenarios link the activity to accretion onto a central SMBH
in one of the two galaxies: 

(i) Accretion disk instability: If an accretion
disk is present in the system, the 
accretion
rate would have to be rather low, 
since there is no
multi-wavelength evidence for an AGN. 
The disk may then settle into the advection dominated (ADAF) mode. 
Whereas ADAFs in general would be characterized by a larger 
extent of the X-ray emitting region as compared to AGN, implying  
less variability than usual (e.g., Ptak et al. 1998), a localized instability in such
a disk could still produce an outburst.
Also, the disk would be more compact in a Kerr metric. 

The thermal instability of slim accretion disks was
studied by, e.g., Honma et al. (1991).
They find the disk to 
exhibit burst-like oscillations for the case of the standard $\alpha$
viscosity description and for certain values of accretion rate.
However, in case of repeated such outbursts in RX\,J1242-11, one might still expect 
to see permanent AGN-typical NLR emission lines (like [OII], [OIII], and [SII])
in the optical spectrum of this galaxy 
(in contrast to emission from what is known
as the BLR in Seyferts; these clouds would have much higher densities
than the NLR and a correspondingly lower recombination timescale.)   

(ii) Tidal disruption of a star:  
Depending on its trajectory, a star gets tidally disrupted after passing a
certain distance to the black hole, the tidal radius, and
the debris is accreted by the hole.
This produces a flare of electromagnetic radiation, lasting on the order
of  months (Rees 1988, 1990).
The peak luminosity should be a substantial fraction of the Eddington 
luminosity.
Rees pointed out that these flares would be excellent indicators
of the presence of SMBHs in nearby, {\em non-active}
galaxies.
The X-ray outburst of RX\,J1242-11 might have originated from
such a tidal disruption event. 
We can then roughly estimate a lower limit on the SMBH mass in RXJ\,1242-11
via the Eddington luminosity which leads to 
$M_{\rm BH} \approxgt 7\,10^{5}$ M$_{\odot}$ if we use our conservative
lower limit on the outburst luminosity. 
More detailed comparisons, e.g., in terms of spectral fits
based on this scenario would have to await more sophisticated
model calculations. In particular, the flares cannot be standardised
and observations will depend  on many parameters, like the type of disrupted star, the impact
parameter, the spin of the black hole, effects of relativistic precession,
 and the radiative transfer is complicated
by effects of viscosity and shocks (Rees 1990).  

(iii) A more speculative scenario would link the outburst to
some merger-induced onset of fuelling of the central region
(e.g, Mihos, 1999). 
There are clear signs that the galaxy pair is interacting{\footnote{
Since many mergers are known to be IR-luminous, we checked the
IRAS faint source catalogue. The two galaxies are not listed
as IR sources, though.}}. 
Athough their distance is presently quite large (about 15 kpc), they
may have already had a closer encounter and the velocity
field in the central region might be disturbed, favoring
accretion events onto the SMBH. 

Follow-up X-ray observations, and long-term monitoring 
if the source turns out to be still ``on"
would provide valuable further clues on the nature of this peculiar
source.

\section{Summarizing conclusions}

We have reported the detection of a giant X-ray outburst from
the previously unknown galaxy pair RX\,J1242.6-1119. 
The outburst X-ray spectrum is very soft (photon index \G $\simeq -5$)
and luminous ($L_{\rm x} > 9\,10^{43}$ erg\,s$^{-1}$), whereas
the optical spectra of both galaxies do not show evidence for Seyfert activity.  
We therefore suggest that  
RX\,J1242-11 is another of the rare cases of giant UV/X-ray outbursts
from non-active galaxies.
The X-ray variability is most likely linked to 
an accretion event -- e.g., by the tidal disruption of a star
as predicted by Rees (1988) --
 on a SMBH residing in the center of
one of the two galaxies.  

Such X-ray outbursts thus provide important information
on the presence of SMBHs in non-active galaxies,
the accretion history of the universe, and the link
between active and normal galaxies.
Further exploitation of the \ros data base 
and future X-ray surveys (like the one that was planned
with {\sl ABRIXAS})  will
be very valuable in finding further of these outstanding  
sources.

\begin{acknowledgements}
It is a pleasure to thank Joachim Tr\"umper, Hartmut Schulz, 
Dirk Grupe, Jules Halpern, Norbert Schartel, Vadim Burwitz, Thomas Boller,  
and Stefan Immler 
for a critical
reading of the manuscript and/or for fruitful discussions,
and the referee, Jane Turner, for her many useful comments. 
The \ros project was supported by the German Bundes\-mini\-ste\-rium
f\"ur Bildung und Forschung
(BMBF/DLR) and the Max-Planck-Society.
JG is supported by
BMBF/DLR under contract No.
FKZ 50 QQ 9602 3. This and related papers can be retrieved from our homepage at
http://www.xray.mpe.mpg.de/$\sim$skomossa/
\end{acknowledgements}

\end{document}